\documentclass[prx,noshowpacs,english,superscriptaddress,twocolumn]{revtex4}
\usepackage{amsmath,amssymb,amsfonts,mathrsfs}
\usepackage{amsthm}
\usepackage{CJK}
\usepackage{bm}
\usepackage{babel}
\usepackage{graphicx}
\allowdisplaybreaks
\usepackage{braket}
\usepackage[breaklinks]{hyperref}
\hypersetup{colorlinks=true, linkcolor=blue, citecolor=blue, filecolor=blue, urlcolor=blue}
\usepackage{booktabs}
\usepackage{multirow}
\usepackage{longtable}

\begin{document}

\title{Triple hourglass Weyl phonons}

\author{Guang Liu}
\affiliation{Department of Physics, Southern University of Science and Technology,
Shenzhen 518055, People's Republic of China}

\author{Zhongjia Chen}
\affiliation{Department of Physics, Southern University of Science and Technology,
Shenzhen 518055, People's Republic of China}
\affiliation{Songshan Lake Materials Laboratory, Dongguan, Guangdong 523808, People's Republic of China}

\author{Peng Wu}
\affiliation{Department of Physics, Southern University of Science and Technology,
Shenzhen 518055, People's Republic of China}

\author{Hu Xu}
\email{xuh@sustech.edu.cn}
\affiliation{Department of Physics, Southern University of Science and Technology,
Shenzhen 518055, People's Republic of China}

\begin{abstract}
Unconventional Weyl phonons with higher topological charges in crystalline solids have attracted increasing attention.
By symmetry analysis and low-energy $\emph{\textbf{k}}\cdot \emph{\textbf{p}}$ effective Hamiltonian,
we propose the symmetry enforced triple hourglass Weyl phonons (THWPs) with Chern number $\mathcal C$ = $\pm$ 3 protected by $6_3$ screw rotation symmetry in chiral space groups 173 ($P6_3$) and 182 ($P6_322$).
We take LiIO$_3$ with space group 173 as a candidate and confirm that it possesses THWP with linear and quadratic dispersions along the $k_z$ direction and in the $k_x$-$k_y$ plane, respectively. Due to the constraints of crystal symmetry and topological charge conservation,
six equivalent single Weyl phonons (SWPs) emerge and lie in the $k_z$ = 0 plane.
Therefore, the unique phonon surface arcs connect the projections of two THWPs and six SWPs, leading to nontrivial sextuple-helicoid surface arcs on the (001) surface
Brillouin zone. Our work proposes a class of topological phonons and realizes it in realistic materials, providing a perfect platform for experimental observation of THWPs. We expect our work to provide a new idea for detection of unconventional quasiparticles.
\end{abstract}



\maketitle

\makeatletter
\def\@hangfrom@section#1#2#3{\@hangfrom{#1#2#3}}
\makeatother

In recent years, various topological quasiparticles in three-dimensional (3D) crystalline solids,
such as Weyl points\cite{PhysRevLett.107.127205,PhysRevLett.108.266802,Yuan2018},
triple points\cite{Park2021,PhysRevLett.120.016401,PhysRevLett.121.035302}, Dirac points\cite{PhysRevLett.108.140405,science.1245085,
RevModPhys.90.015001,Cai2020,PhysRevLett.126.185301},
nodal lines\cite{PhysRevLett.115.036806,PhysRevLett.115.036807,PhysRevB.98.220103,PhysRevB.104.024304},
and nodal surfaces\cite{PhysRevB.93.085427,PhysRevB.97.115125,PhysRevB.104.L041104}, etc.,
have attracted widespread attention because of their unique physical properties and potential applications.
Among them, Weyl-type excitations are of particular importance, which are peculiar quasiparticle excitations featuring
nontrivial isolated point touching of two band branches.  Weyl fermions were first predicted theoretically
in the magnetic pyrochlore iridates\cite{PhysRevB.83.205101}.
However, due to the limitation
of magnetic domains of these magnetic materials, some typical physical properties of Weyl points are difficult to be observed experimentally\cite{PhysRevB.86.115133,PhysRevB.88.104412,PhysRevX.5.031023}.
Fortunately, non-magnetic materials with well-defined surface states and Fermi arcs
are later proposed and observed experimentally\cite{PhysRevX.5.011029,PhysRevX.5.031013,science.aaa9273,science.aaa9297,Lv2015}.

\begin{figure*}
\centering
\renewcommand{\figurename}{Fig.}
\setlength{\abovecaptionskip}{-2.3 cm}
\includegraphics[scale=0.4]{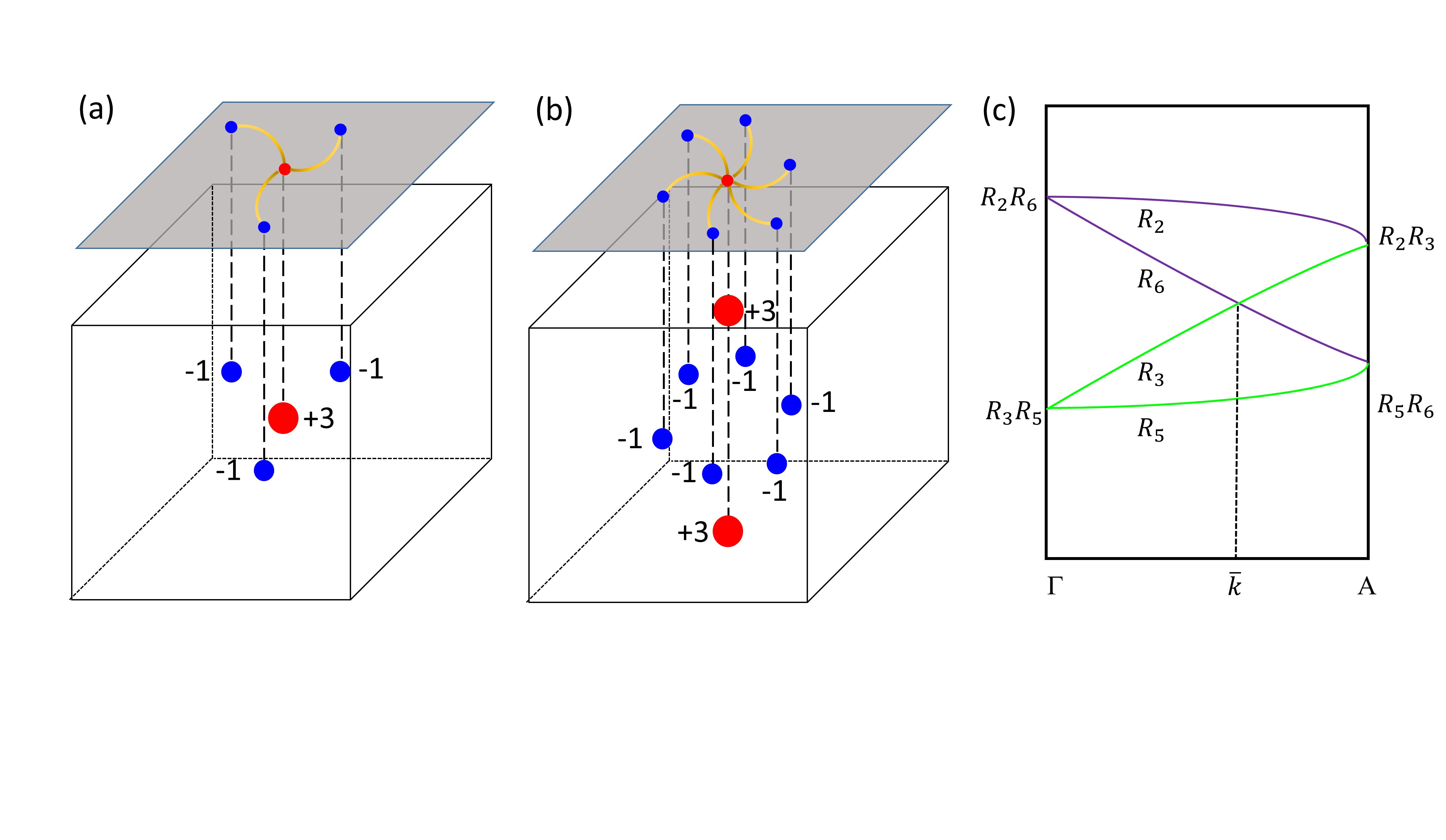}
\caption{(a) The schematics of a triple Weyl point with triple-helicoid surface arcs. (b) The schematic of a pair of triple Weyl points connected by time-reversal symmetry with sextuple-helicoid surface arcs. (c) The schematic of the hourglass band dispersion on  a $6_3$ invariant screw axis.}
\end{figure*}


Motivated by these works, more works on Weyl-type excitations have been reported.
According to the band dispersion around crossing points which can lead to exotic topological properties, the Weyl-type points
exhibit a variety of classification.
For example, according to degree of the tilt of the local dispersion around the crossing,
Weyl points are divided into type-I\cite{Yuan2018,PhysRevB.83.205101,PhysRevX.5.011029,PhysRevX.5.031013,science.aaa9273,
science.aaa9297,Lv2015,PhysRevLett.122.057205}, type-II\cite{Soluyanov2015,Deng2016,PhysRevX.6.031021,
PhysRevLett.117.077202,PhysRevLett.123.065501,PhysRevB.105.035141}, and type-III Weyl
points\cite{PhysRevB.101.100303,PhysRevB.103.L081402,PhysRevB.105.134303}.
The latter two break the Lorentz symmetry, resulting in
unique transport properties different from type-I Weyl points with normal point-like
Fermi surfaces\cite{Soluyanov2015,Deng2016,PhysRevX.6.031021,
PhysRevLett.117.077202,PhysRevLett.123.065501,PhysRevB.105.035141}.
In addition, the Weyl points, which act as ``source" or ``sink" of Berry curvature\cite{science.1089408},
are topologically protected by the discrete translational symmetry of the lattice and no further symmetries
are needed for their existence. Therefore, the topology of Weyl point can
be defined by a quantized topological charge (i.e., the Chern number $\mathcal C$ = $\pm$1), which is the flux of the Berry
curvature passing through a closed surface in momentum space.
However, considering an additional symmetry, there
will exist some Weyl points with higher Chern number $\mathcal     C$\cite{PhysRevB.96.045102,PhysRevB.95.125136,PhysRevLett.124.105303,Huang2020,YU2022375,PhysRevB.103.104101,PhysRevB.102.125148,PhysRevB.103.184301,PhysRevB.104.075115,PhysRevB.103.L161303}. For example, a fourfold or sixfold rotation symmetry can protects Weyl points with
$\mathcal C$=$\pm2$\cite{PhysRevLett.120.016401,PhysRevB.96.045102,PhysRevB.95.125136,
PhysRevLett.124.105303,Huang2020}, a sixfold rotation symmetry can protects Weyl points with $\mathcal C$=$\pm3$\cite{PhysRevLett.108.266802,YU2022375},
and Weyl points with $\mathcal C$=$\pm4$\cite{PhysRevB.102.125148,PhysRevB.103.184301,PhysRevB.104.075115,PhysRevB.103.L161303} can be realized when the cubic symmetry is included in the presence of threefold screw rotational symmetry. Under this convention, Weyl points are divided into single Weyl, quadratic Weyl, triple Weyl, and charge-four Weyl points with
$\mathcal C$=$\pm1$, $\pm2$, $\pm3$, and $\pm4$, respectively.

Phonons can be a perfect platform for realization of these unconventional Weyl points
due to their particular device applications and the unique advantages of whole frequency range observation. For example,
quadratic Weyl phonons (QWPs) can exist with other two equal single Weyl phonons (SWPs) in $\alpha$-SiO$_2$\cite{PhysRevLett.124.105303}.
Unlike paired  SWPs with opposite chiral charges, the projection of QWP
connects two SWPs by two surface arcs.
More importantly, such a combination does not violate the no-go theorem\cite{NIELSEN198120,NIELSEN1981173} and satisfies topological charge conservation.
The charge-four Weyl (CFW) point was proposed only in the spinless system\cite{PhysRevB.102.125148,PhysRevB.104.075115,PhysRevB.103.L161303},
and the quadruple-helicoid surface states are available.
However, the studies focusing on topological nontrivial triple Weyl phonons (TWPs) are still in its infancy,
and the TWPs with clearly visible surface states have so far not been realized in realistic materials. As shown in Fig. 1(a),
we propose the existence of the TWP with exotic triple-helicoid surface states, which connect the projections of the TWP and three SWPs.
Interestingly, sextuple-helicoid surface states in Fig. 1(b) can be expected when two TWPs are connected by time-reversal symmetry $\mathcal T$.
However, the TWPs originate from the accidental degeneracies on a C$_6$ invariant line, and therefore we can not
explicitly distinguish it from SWPs and QWPs on the same path.
Then a questions arises: \emph{Is there some kind of constraint that guarantees the existence of TWPs on high-symmetry lines (HSLs)?}
The hourglass-shaped band structures, which may result in symmetry enforced intersection, such as hourglass nodal points and nodal lines\cite{PhysRevB.96.075110,FURUSAKI2017788,Wang2017,PhysRevB.104.L060301,PhysRevMaterials.6.034202}, would be a
good choice to realize the TWPs. Therefore, a class of exotic topological phase, i.e., the triple hourglass Weyl phonons (TWHPs), can be expected.

Through $\emph{\textbf{k}}\cdot \emph{\textbf{p}}$ effective model analysis, we identified all candidate space groups (SGs) capable of possessing TWPs, which are listed in Table. S1 of the supplementary material (SM)\cite{SM}. On this basis, by analyzing the evolution of the irreducible representations (IRs) on C$_6$
invariant HSLs, we find that SGs 173 and 182 can
host the THWPs with $\mathcal C$ = $\pm$3 in our candidate SGs, which originates from the
different screw rotation degrees of the C$_6$. Therefore, we divide the triple Weyl phonons into general TWPs  and THWPs.
Taking LiIO$_3$ as an example, we confirm the nontrivial topological properties of THWPs
and find it possesses an unique sextuple-helicoid surface arcs on the (001) surface, which is consistent with our prediction.
It should be stressed that THWP are enforced to appear by symmetry,
and its hourglass band characteristics makes it easier to be observed and detected.

To understand the properties of TWPs, the crossings of two phonon branches
can be described by a two-band $\emph{\textbf{k}}\cdot \emph{\textbf{p}}$ effective Hamiltonian, namely
\begin{equation}
\mathcal H(\emph{\textbf{k}})=d(\emph{\textbf{k}})\sigma_++d^*(\emph{\textbf{k}})\sigma_-+f(\emph{\textbf{k}})\sigma_z
\end{equation}
where $\sigma_{x,y,z}$ are Pauli matrix, $\sigma_\pm=(\sigma_x \pm i\sigma_y)/2$,  $d(\emph{\textbf{k}})$ represents a
complex function, and $f(\emph{\textbf{k}})$ represents a real function.
As an illustration, we consider a generic point $k$ on the $C_6$
invariant axis [$\Gamma$-Z $(0,0,w)$] of hexagonal crystal system. Since
[$C_{6z}$, $\mathcal H(\emph{\textbf{k}})$] = 0, the band eigenstates at $\emph{\textbf{k}}$ can be simultaneously chosen as $C_{6z}$ eigenstates. The little
group can be generated by the screw rotation symmetry $\widetilde{C}_{6z}\equiv\{C_{6z}|00\frac{u}{6} \}$, changing the
coordinates of the lattice as

\begin{equation}
\begin{aligned}
\widetilde{C}_{6z}: (x, y, z)\rightarrow(\frac{1}{2}x-\frac{\sqrt{3}}{2}y,\frac{\sqrt{3}}{2}x+\frac{1}{2}y, z+\frac{u}{6}),\\
\end{aligned}
\end{equation}
where $u$ = 0, 1, 2, 3, 4, 5. Considering a general point located at the $k_z$ axis (0,0,$w$). Then we can get the relation
\begin{equation}
\widetilde{C}_{6z}^6=T_{00u}=e^{i2u\pi w}.
\end{equation}
Hence, the eigenvalues of $\widetilde{C}_{6z}$ can be expressed as
\begin{equation}
E_n=\{e^{in\pi /6}e^{i\pi uw/3},n=0,2,4,6,8,10\},
\end{equation}
where $E_n$ can be used to indicate
each phonon branch. When two phonon branches with different eigenvalues cross on this $\widetilde{C}_{6z}$ invariant
path, the matrix representation of $\widetilde{C}_{6z}$ can be written as
\begin{equation}
D(\widetilde{C}_{6z})=\left(
\begin{array}{cc}
E_{n_1}  & \\
&E_{n_2}\\
\end{array}
\right),
\end{equation}
where $n_1$/$n_2$ represents different $n$ in Eq. (4). According to the invariant theory, the Hamiltonian of this crossing point can be expressed under the constraint of $\widetilde{C}_{6z}$ invariant symmetry as
\begin{equation}
D(\widetilde{C}_{6z})\mathcal{H}_{eff}(\emph{\textbf{k}})D^{-1}(\widetilde{C}_{6z})=\mathcal{H}_{eff}(R_{6z}\emph{\textbf{k}}),
\end{equation}
where $R_{6z}$ is a 3$\times$3 rotation matrix of $\widetilde{C}_{6z}$.
There exist three possibilities, leading to three types of Weyl point:
 \begin{equation}
E_{n_1}/E_{n_2}=e^{\pm i\pi /3}, \mathcal C=\pm1,
\end{equation}
\begin{equation}
E_{n_1}/E_{n_2}=e^{\pm 2i\pi /3}, \mathcal C=\pm2,
\end{equation}
\begin{equation}
E_{n_1}/E_{n_2}=-1, \mathcal C=\pm3.
\end{equation}
Under the constraint of Eqs. (6) and (9), the crossing point would be a triple Weyl point, and the corresponding Hamiltonian can be expressed as
\begin{equation}
\mathcal{H}_{eff}(\emph{\textbf{k}})=\left(
\begin{array}{cc}
f(\emph{\textbf{k}}) &d(\emph{\textbf{k}}) \\
d^*(\emph{\textbf{k}}) & -f(\emph{\textbf{k}})\\
\end{array}
\right),
\end{equation}
where $f(\emph{\textbf{k}})$ = $a_1(k_x^2+k_y^2)+a_2k_z$, $d(\emph{\textbf{k}})$ = $\alpha_1k_+^3+\alpha_2k_-^3$,
$a_{1,2}$ are real parameters, $\alpha_{1,2}$ are complex parameters, and $k_\pm=k_x \pm ik_y$.
In this low-energy effective Hamiltonian, the topological charge of the crossing point is characterized
by the leading order of the chiral term [i.e., the antidiagonal term].
Besides, the in-plane cubic dispersion originating from the the chiral term is suppressed
by the in-plane quadratic dispersion of the nonchiral term [i.e., diagonal term],
so the triple Weyl point possesses the in-plane [$k_x$-$k_y$] quadratic dispersion in a spinless system.

\begin{table*}
\centering
\setlength{\tabcolsep}{2.5mm}
\caption{IRs of $C_6$ with different screw rotation degrees at high symmetry points (HSPs) $\Gamma$ and A, where $u$ represents screw rotation degree. $R_1$, $R_2$, $R_3$, $R_4$, $R_5$, and $R_6$ represent different IRs corresponding to $n$ = 0, 2, 4, 6, 8, and 10 in Eq. (4), respectively. }
\begin{tabular}{c|c|cccccc|c|c}
\hline \hline
\ Degree & HSP   & $R_1$ & $R_2$ & $R_3$ & $R_4$ & $R_5$ & $R_6$ & 2D IRs&Triple Weyl on $\Gamma$-A\\

\hline
\ \multirow{2}{*}{$u$=0} & $\Gamma$  &  1                    & $e^{i\pi/3}$       & $e^{2i\pi/3}$       &-1            &$e^{-2i\pi/3}$      &$e^{-i\pi/3}$ &$R_{2,6}$,$R_{3,5}$& \multirow{2}{*}{$R_1R_4/R_2R_5/R_3R_6$}\\
\ & A                &  1                    & $e^{i\pi/3}$       & $e^{2i\pi/3}$       &-1            &$e^{-2i\pi/3}$      &$e^{-i\pi/3}$ &$R_{2,6}$,$R_{3,5}$& \\

\hline
\ \multirow{2}{*}{$u$=1} &  $\Gamma$  &  1                    & $e^{i\pi/3}$       & $e^{2i\pi/3}$       &-1            &$e^{-2i\pi/3}$    &$e^{-i\pi/3}$&$R_{2,6}$,$R_{3,5}$&\multirow{2}{*}{$R_1R_4/R_2R_5/R_3R_6$}  \\
\ &  A                &$e^{i\pi/6}$                   & i   &$e^{5i\pi/6}$       &$e^{-5i\pi/6}$               &-i      & $e^{-i\pi/6}$&$R_{1,6}$,$R_{2,5}$ ,$R_{3,4}$& \\

\hline
\ \multirow{2}{*}{$u$=2} &  $\Gamma$  &  1                    & $e^{i\pi/3}$       & $e^{2i\pi/3}$       &-1            &$e^{-2i\pi/3}$      &$e^{-i\pi/3}$ &$R_{2,6}$,$R_{3,5}$&\multirow{2}{*}{$R_1R_4/R_2R_5/R_3R_6$}  \\
\  & A                &  $e^{i\pi/3}$                    & $e^{2i\pi/3}$    &-1       & $e^{-2i\pi/3}$              & $e^{-i\pi/3}$       &   1  &$R_{1,5}$,$R_{2,4}$& \\

\hline
\  \multirow{2}{*}{$u$=3} & $\Gamma$  &  1                    & $e^{i\pi/3}$       & $e^{2i\pi/3}$       &-1            &$e^{-2i\pi/3}$      &$e^{-i\pi/3}$ &$R_{2,6}$,$R_{3,5}$&\multirow{2}{*}{$R_1R_4/R_2R_5/R_3R_6$}  \\
\ &  A                &   i                    & $e^{5i\pi/6}$    &$e^{-5i\pi/6}$       &-i              &$e^{-i\pi/6}$       & $e^{i\pi/6}$ &$R_{1,4}$,$R_{2,3}$,$R_{5,6}$ &\\

\hline
\ \multirow{2}{*}{$u$=4} &  $\Gamma$  &  1                    & $e^{i\pi/3}$       & $e^{2i\pi/3}$       &-1            &$e^{-2i\pi/3}$      &$e^{-i\pi/3}$&$R_{2,6}$,$R_{3,5}$ &\multirow{2}{*}{$R_1R_4/R_2R_5/R_3R_6$} \\
\  & A                & $e^{2i\pi/3}$               & -1    &$e^{-2i\pi/3}$       &$e^{-i\pi/3}$              & 1       & $e^{i\pi/3}$&$R_{1,3}$,$R_{4,6}$ & \\

\hline
\  \multirow{2}{*}{$u$=5} & $\Gamma$  &  1                    & $e^{i\pi/3}$        & $e^{2i\pi/3}$       &-1                     &$e^{-2i\pi/3}$      &$e^{-i\pi/3}$ &$R_{2,6}$,$R_{3,5}$&\multirow{2}{*}{$R_1R_4/R_2R_5/R_3R_6$} \\
\ &  A                &  $e^{5i\pi/6}$ & $e^{-5i\pi/6}$     &-i                          &$e^{-i\pi/6}$     &$e^{i\pi/6}$        & i                   &$R_{1,2}$,$R_{3,6}$,$R_{4,5}$&  \\

\hline \hline
\end{tabular}
\end{table*}

Based on the symmetry analysis above, we list the operation eigenvalues of the C$_6$ screw rotation operation at high symmetry points (HSPs) $\Gamma$ and A in Table I.
Under the action of time-reversal symmetry, two one-dimensional (1D) IRs can form a two-dimensional (2D) IR. For example, when $u$ = 0 at $\Gamma$, $R_{2,6}= R_2\oplus R_6$. In Table I, we list all 2D IRs at high symmetry $\Gamma$ and A points, respectively.
Then we consider the evolution of two sets of 2D IRs along the HSL $\Gamma$-A. The two sets of 2D IRs will split into four 1D IRs
along $\Gamma$-A. When the screw rotation $6_3$ [i.e., $\widetilde{C}_{6z}$ with $u$ = 3] is presented, interesting things happen.
As shown in Fig. 1(c), two sets of 2D IRs $R_{2,6}$ and $R_{3,5}$ at $\Gamma$ split into four 1D IRs $R_2$,  $R_6$, $R_3$, and $R_5$ along $\Gamma$-A, then form two 2D IRs
$R_{2,3}$ and $R_{5,6}$ at A. According to the compatibility relations, there must a crossing with $\mathcal C$ = $\pm3$ between
$R_3$ and $R_6$. In contrast to the general triple Weyl phonon,
this crossing is enforced by $6_3$ screw rotation symmetry and has a hourglass type band dispersion structure, which is the THWP. Interestingly, the topological charge of the THWP can be obtained explicitly from the band dispersion without further calculations.
In addition, we find that only SGs 173 and 182 satisfy such a symmetry requirement.

\begin{figure}
\centering
\renewcommand{\figurename}{Fig.}
\setlength{\abovecaptionskip}{-0.3 cm}
\includegraphics[scale=0.30]{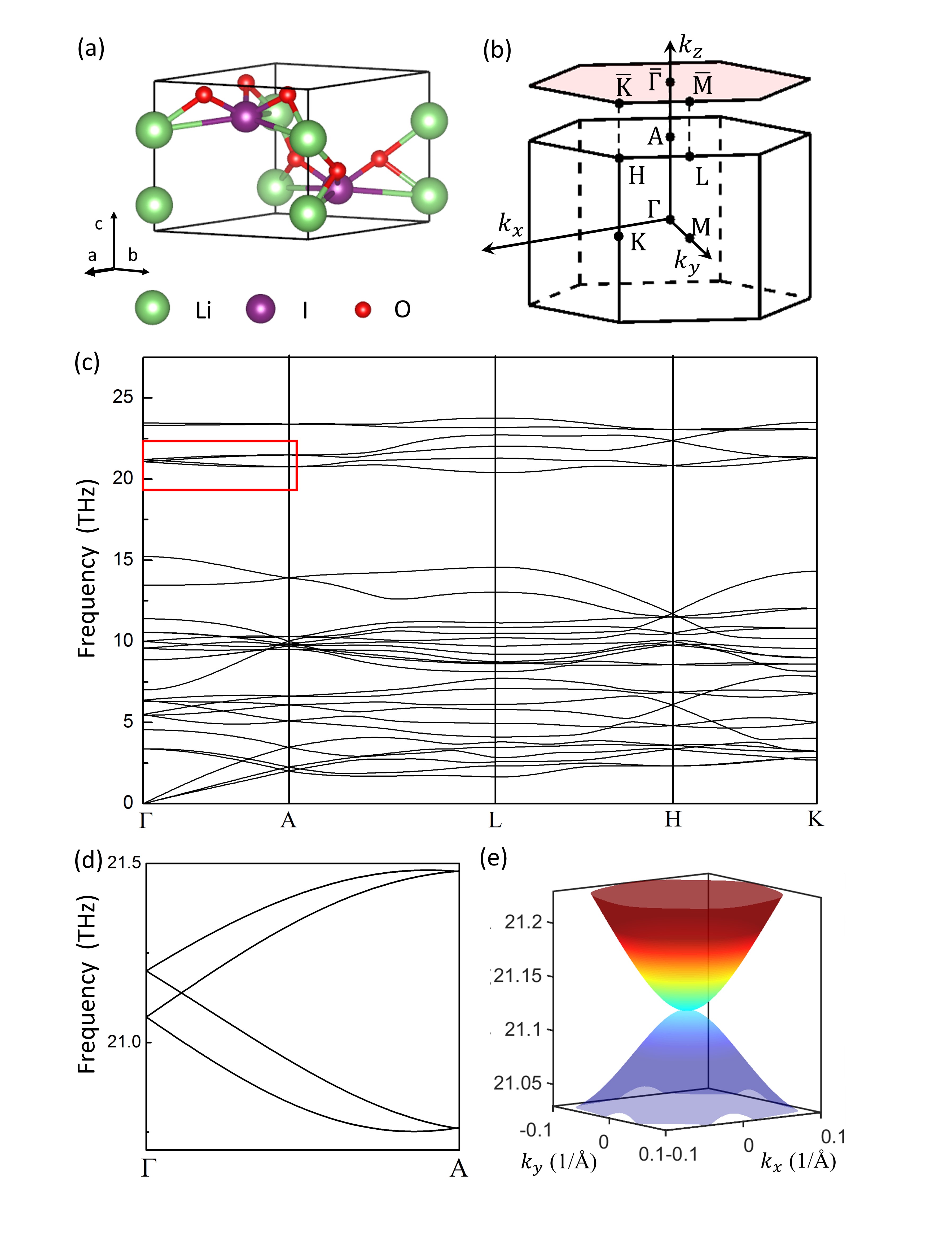}
\caption{The crystal structure and phonon band of LiIO$_3$.(a) The primitive cell.
(b) The bulk BZ and the corresponding (001) surface BZ. (c) Phonon spectrum. (d) Phonon
dispersions along $\Gamma$-A contributed from phonon branches 25 to 28, which corresponding to the red box in (c). (e) Three-dimensional representation of the phonon dispersion
on the $k_x$-$k_y$ plane of the crossing point in (d). }
\end{figure}

Lithium carbonate ($\alpha$-LiIO$_3$) is an ideal candidate to realize the symmetry enforced THWPs,
which is synthesized by reagent grade iodic acid and lithium carbonate dissolved in water\cite{a05127}. As shown in Fig. 2(a), the SG of
$\alpha$-LiIO$_3$ is 173 ($P6_3$), in which a primitive cell contains two Li, two I, and six O atoms. The bulk Brillouin zone (BZ),  the projected
(001) surface BZ of $\alpha$-LiIO$_3$ and the corresponding HSPs are shown in Fig. 2(b). As shown in Fig. 2(c), we plot the phonon band structure with non-analytic term correction for $\alpha$-LiIO$_3$.
We focus on the non-trivial hourglass crossing on the C$_6$ invariant axis, which corresponds to the
THWP. We find that the hourglass crossing contributed from the phonon branches
25 to 28 is an ideal selection. As shown in Fig. 2(d), the THWP has a typical linear dispersion characteristic
along the $k_z$ direction. We also plot the corresponding 3D representation of the crossing branches in the $k_x$-$k_y$ plane.
It is clear that the crossing point possesses quadratic in-plane dispersion in Fig. 2(e), which is consistent with our model analysis.

\begin{figure}
\centering
\renewcommand{\figurename}{Fig.}
\setlength{\abovecaptionskip}{-0.5 cm}
\includegraphics[scale=0.48]{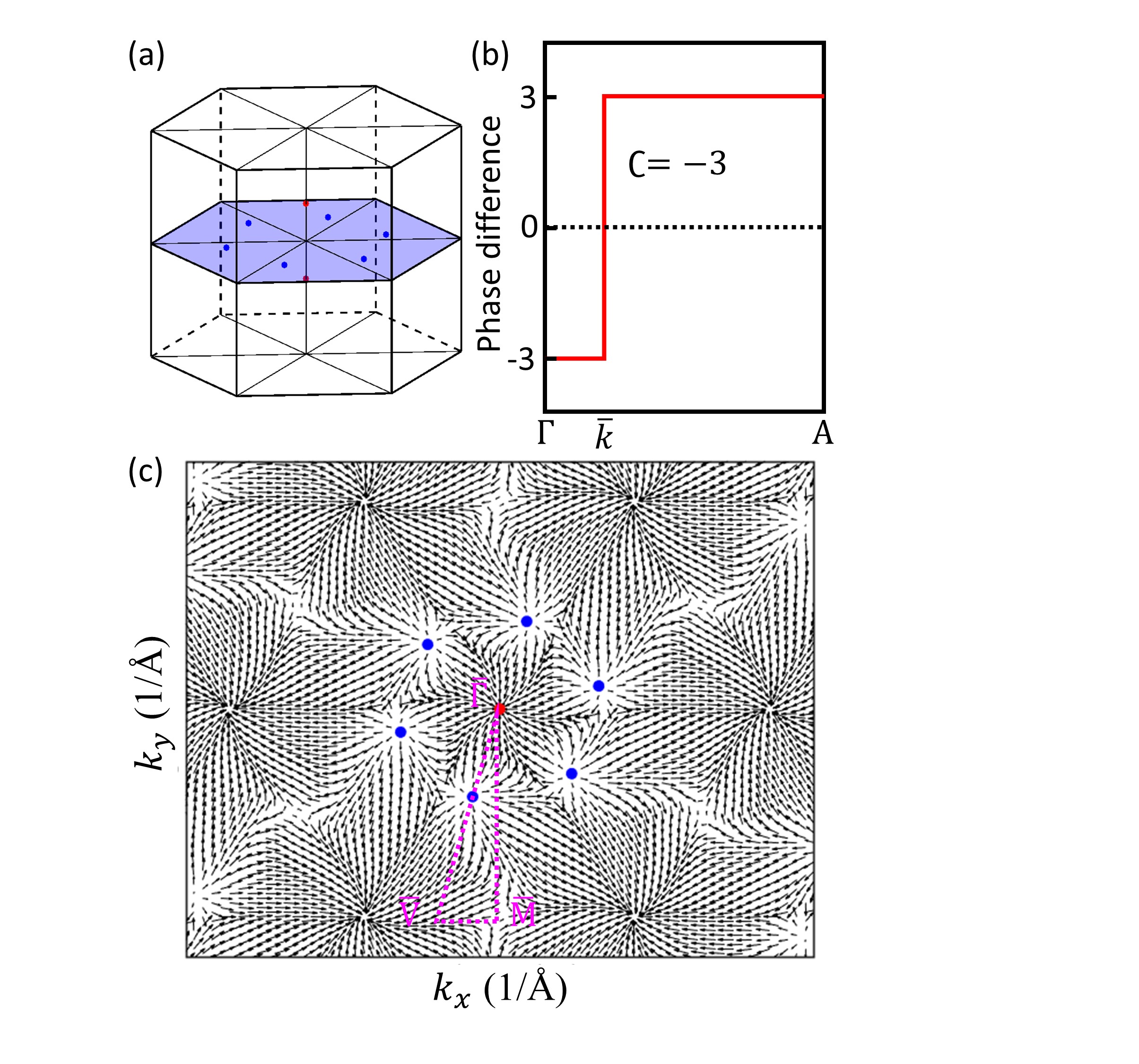}
\caption{(a) The distribution of the THWP (red dots) and SWPs (blue dots) in momentum space.
(b) The phase difference evolution of $C_6$ rotation operation eigenvalues contributed by  two phonon branches 26 and 27 along $\Gamma$-A.
(c) The distribution of Berry curvature in the $k_x$-$k_y$ plane. $\overline {\Gamma}$ - $\overline {\textrm M}$ - $\overline {\textrm V}$ represents a (001) surface path through the projections of THWPs and SWPs .}
\end{figure}

We also find six SWPs contributed from 26 and 27 phonon branches at the $k_z$ = 0 plane,
and the schematic diagram about the momentum positions of two THWPs and six SWPs in the first BZ are shown in Fig. 3(a).
Under the rotation of $C_6$ symmetry operator, these six SWPs are equivalent, and first-principles calculations show
that their Chern number are +1. According to the no-go theorem, we can also infer that the Chern numbers of the two THWPs along
the $k_z$-axis are -3. The momentum positions and more detailed information about these
nodes are presented in Table S2 of the SM.

Furthermore, in order to explicitly obtain the chirality of THWP, we study the phase difference [i.e., $3/ \pi* arg(E_{n_1}/E_{n_2})$] evolution of the $C_6$ eigenvalues of two phonon branches in Fig. 3(b). We find that there exists a phase change at the $\overline k$ point, confirming the existence of a THWP with $\mathcal C$ = -3.
We next examine the distribution of Berry curvature
contributed by the 26th and 27th phonon branches in the $k_x$-$k_y$ plane. As shown in Fig. 3(c), the Berry curvature near
topological nontrivial points shows convergent and divergent field morphology, which correspond to the chirality
of THWPs and SWPs, respectively.

\begin{figure}
\centering
\renewcommand{\figurename}{Fig.}
\setlength{\abovecaptionskip}{-1 cm}
\includegraphics[scale=0.4]{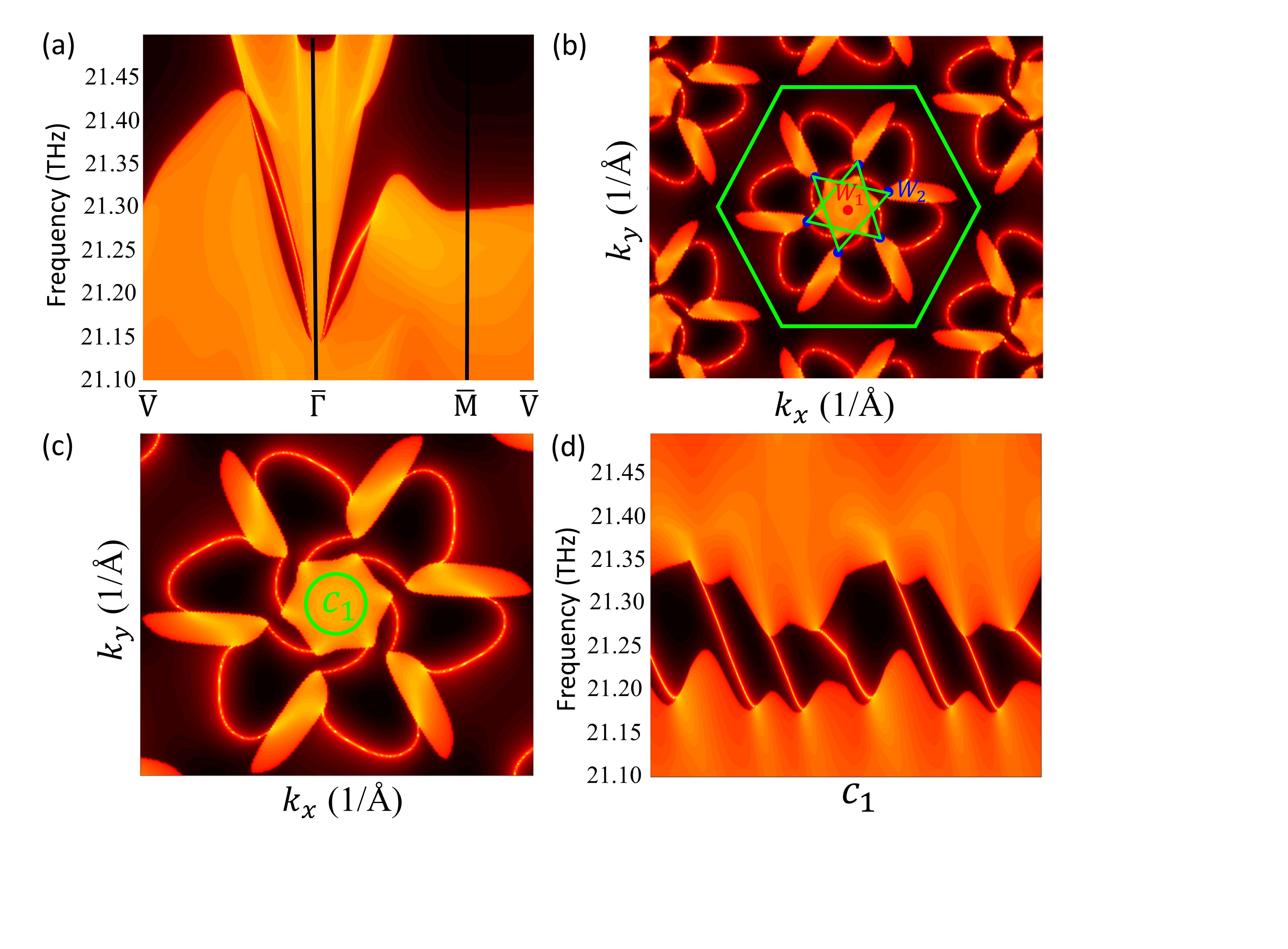}
\caption{(a)The phonon LDOS projected on the (001) surface BZ along
$\overline {\textrm V}$ - $\overline {\Gamma}$ - $\overline {\textrm M}$ - $\overline {\textrm V}$.
(b) The isofrequency contours of (001) surface and (c) its corresponding enlarged view of the middle part at 21.37 THz, where $W_1$ and $W_2$ represent the projection of THWPs and SWPs, respectively. (d) The phonon LDOS projected
on the semi-infinite (001) surface BZ along the loop $C_1$.  }
\end{figure}

To illustrate the exotic topological features of THWPs, we calculate the phonon local density of states
(LDOS) and projected isofrequency surface contours on the (001) surface.
The phonon surface states, which start at the projection of the THWPs
at $\overline \Gamma$ and end at the projection of SWPs locating at $\overline \Gamma$-$\overline{\textrm V}$, are clearly visible in Fig. 4(a).
Moreover, the (001) surface states further display quadratic dispersions of the THWPs,
showing consistency with the 3D phonon dispersion in Fig. 2(e). As shown in Fig. 4(b), two THWPs  located at $\Gamma$-$\textrm A$
are projected into the center of the surface BZ [denoted by $W_1$], forming a terminal point with total $\mathcal C$ = -6, while the other six
SWPs are projected into the internal general point of the surface BZ [denoted by $W_2$] with $\mathcal C$ = +1. We can further determine that the
six non-trivial terminal points corresponding to six SWPs in the isofrequency surface contours
with $\omega$ = 21.37 THz in Fig. 4(b).
And we can see that six branches of surface arcs start at $W_2$, and finally converge at $W_1$. These long surface arcs that connect $W_1$ and $W_2$
exhibit a sextuple-helicoid nature. Unlike the QWPs that
usually accompanies the double-helicoid surface state, each THWP induces the simultaneous appearance of three
other SWPs, forming the triple-helicoid surface arc structure. The THWPs
can only appear on the $C_6$ invariant axis, and its (001) projection is a
superposition of two THWPs. Therefore, we can see sextuple-helicoid surface states, which is in perfect
agreement with our theoretical prediction. To visualize the helicoidal surface states, we calculate the surface LDOS
along one clockwise loops $C_1$ [see Fig. 4(c)] in Fig. 4(d). For the loop $C_1$, the six leftmoving
chiral edge modes are consistent with the chirality of $W_1$, further confirming our results.

In summary, we present a class of symmetry enforced THWP with unique topological properties, which is protected by $6_3$ screw rotation symmetry in chiral SGs 173 and 182. Our work uncovers all THWPs in the spinless system and proposes that their nontrivial surface states connect
a THWP with $\mathcal C= \pm3$ and three SWPs with $\mathcal C=\mp1$.  In addition, we provide an ideal
candidate which possesses two THWPs locating at $C_6$ invariant line and six SWPs locating at the $k_z=0$ plane, leading to sextuple-helicoid surface arcs. The multiple surface arcs can provide one-way phonon propagation channels, suggesting potential applications in topological phonon devices.
Our work not only proposes a class of quasiparticles but also provides an idea for detection of unconventional quasiparticles in realistic materials.

This work is supported by the National Key R\&D Program of China (Grant No. 2022YFA1403700), the National Natural Science Foundation of China (Grant No. 11974160), the Science,
Technology, and Innovation Commission of Shenzhen Municipality (Grant No. RCYX20200714114523069), and the Center for Computational Science and Engineering at Southern University of Science and Technology.

\textit{Note added}. $-$Recently, we became aware of a similar work by Wang \textit{et al.}\cite{wangxiaotian}.


\newpage

\setcounter{figure}{0}
\makeatletter

\makeatother
\renewcommand{\thefigure}{S\arabic{figure}}
\renewcommand{\thetable}{S\Roman{table}}
\renewcommand{\theequation}{S\arabic{equation}}
\begin{center}
	\textbf{
		\large{Supplemental Material for}}
	\vspace{0.2cm}
	
	\textbf{
		\large{
			``Triple Hourglass Weyl phonons" }
	}
\end{center}



\textbf{\large{A. Calculation methods}}
~\\


The calculations for the realistic materials were based on the framework of density functional theory (DFT) \cite{kohn1965self} using the Vienna \emph{ab-initio} Simulation Package (VASP) \cite{PhysRevB.54.11169,kresse1996efficiency}. The generalized gradient approximation (GGA) with the Perdew-Burke-Ernzerhof revised for solids
(PBE-sol) formalism was employed for the exchange-correlation function \cite{PhysRevLett.77.3865}. The projector augmented wave method
was employed to treat core-valence interactions\cite{PhysRevB.59.1758,PhysRevLett.45.566}. For all calculations, the energy and force
convergence criteria were set to be $10^{-7}$ eV and $10^{-3}$ eV/A, respectively. The plane-wave expansion was truncated with a cutoff energy
of 520 eV, and the full Brillouin zone (BZ) was sampled by $8\times8\times8$ Monkhorst-Pack grid. For phonon spectra calculations, we used the PHONOPY code to construct the force constant matrices and generate the symmetry information \cite{TOGO20151}. The finite displacement method was
used to generate interatomic force constants in a $5\times5\times2$ supercell. In order to calculate the phonon surface states, we constricted the tight-binding model Hamiltonian using the WannierTools package \cite{WU2018405} combined with the iterative Green function method \cite{Sancho_1984}.

~\\
\textbf{\large{B. The summary of space groups with triple Weyl phonons}}
~\\


\renewcommand{\thetable}{S1}
\begin{table*}[h]
\centering
\setlength{\tabcolsep}{12mm}
\caption{The complete list of the triple Weyl phonons (TWPs) in 230 SGs. The first columns indicate the space group (SG) number and symbol, and the
second column indicates the corresponding momentum location. The
third column show the corresponding irreducible representations. The fourth column shows the generators of the k line.}
\begin{tabular}{cccc}
\hline \hline
\ SG & Location & IRs & Generator \\
\hline
\ 168($P6$)  & $\Gamma-\textrm{A}$ & $R_1R_4/R_2R_5/R_3R_6$  & $\widetilde{C}_{6z}\equiv\{C_{6z}|000 \}$\\

\ 169($P6_1$)  & $\Gamma-\textrm{A}$ & $R_1R_4/R_2R_5/R_3R_6$  & $\widetilde{C}_{6z}\equiv\{C_{6z}|00\frac{1}{6} \}$ \\

\ 170($P6_5$)  & $\Gamma-\textrm{A}$ & $R_1R_4/R_2R_5/R_3R_6$  & $\widetilde{C}_{6z}\equiv\{C_{6z}|00\frac{5}{6} \}$\\

\ 171($P6_2$)  & $\Gamma-\textrm{A}$ & $R_1R_4/R_2R_5/R_3R_6$  & $\widetilde{C}_{6z}\equiv\{C_{6z}|00\frac{2}{6} \}$ \\

\ 172($P6_4$)  & $\Gamma-\textrm{A}$ & $R_1R_4/R_2R_5/R_3R_6$  & $\widetilde{C}_{6z}\equiv\{C_{6z}|00\frac{4}{6} \}$ \\

\ 173($P6_3$)  & $\Gamma-\textrm{A}$ & $R_1R_4/R_2R_5/R_3R_6$  & $\widetilde{C}_{6z}\equiv\{C_{6z}|00\frac{3}{6} \}$ \\

\ 177($P622$)  & $\Gamma-\textrm{A}$ & $R_1R_4/R_2R_5/R_3R_6$  & $\widetilde{C}_{6z}\equiv\{C_{6z}|000 \}$ \\

\ 178($P6_122$)  & $\Gamma-\textrm{A}$ & $R_1R_4/R_2R_5/R_3R_6$  & $\widetilde{C}_{6z}\equiv\{C_{6z}|00\frac{1}{6} \}$ \\

\ 179($P6_522$)  & $\Gamma-\textrm{A}$ & $R_1R_4/R_2R_5/R_3R_6$  &$\widetilde{C}_{6z}\equiv\{C_{6z}|00\frac{5}{6} \}$ \\

\ 180($P6_222$)  & $\Gamma-\textrm{A}$ & $R_1R_4/R_2R_5/R_3R_6$  & $\widetilde{C}_{6z}\equiv\{C_{6z}|00\frac{2}{6} \}$ \\

\ 181($P6_422$)  & $\Gamma-\textrm{A}$ & $R_1R_4/R_2R_5/R_3R_6$  & $\widetilde{C}_{6z}\equiv\{C_{6z}|00\frac{4}{6} \}$ \\

\ 182($P6_322$)  & $\Gamma-\textrm{A}$ & $R_1R_4/R_2R_5/R_3R_6$  & $\widetilde{C}_{6z}\equiv\{C_{6z}|00\frac{3}{6} \}$ \\

\hline \hline
\end{tabular}
\end{table*}


~\\
\textbf{\large{C. The momentum positions and more detailed information about THWPs and WPs in LiIO$_3$}}
~\\

In this section, Fig. S1 shows the related atom vibration of  LiIO$_3$. Table S2 shows the momentum positions and more detailed information about THWPs and WPs in LiIO$_3$.
Fig. S2 shows the (100) surface states and surface arcs of LiIO$_3$.

\renewcommand{\thefigure}{S\arabic{figure}}
\begin{figure*}
\centering
\renewcommand{\figurename}{Fig.}
\setlength{\abovecaptionskip}{0 cm}
\includegraphics[scale=0.4]{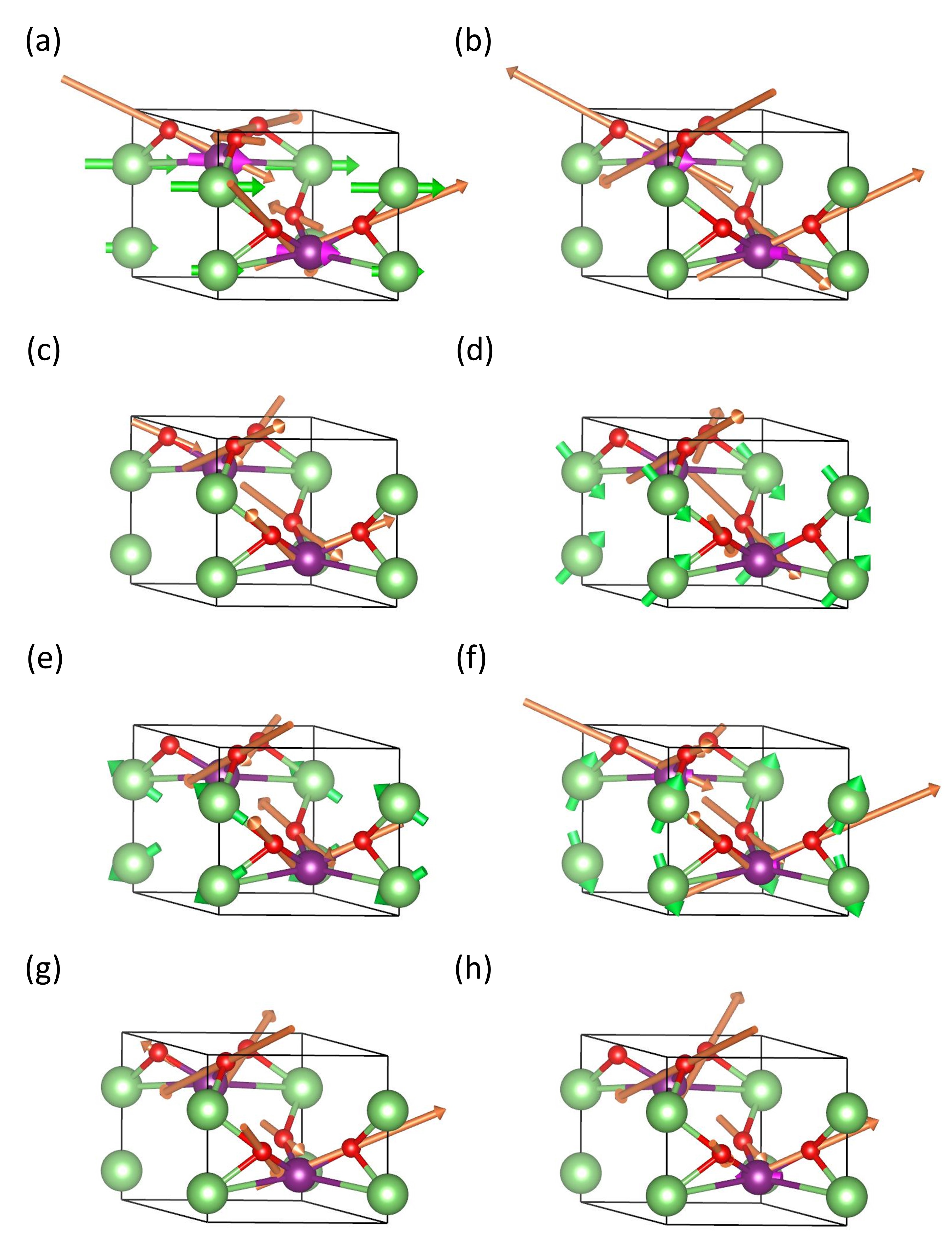}
\caption{The related displacements from the phonon branches 26 (a,c,e,f) and 27 (b,d,f,h) at $\Gamma$, M, K, and A, respectively. The length of the arrow represents the relative magnitude of the vibration.  }
\end{figure*}


\renewcommand{\thetable}{S2}
\begin{figure*}
\centering
\renewcommand{\figurename}{Fig.}
\setlength{\abovecaptionskip}{0 cm}
\includegraphics[scale=0.7]{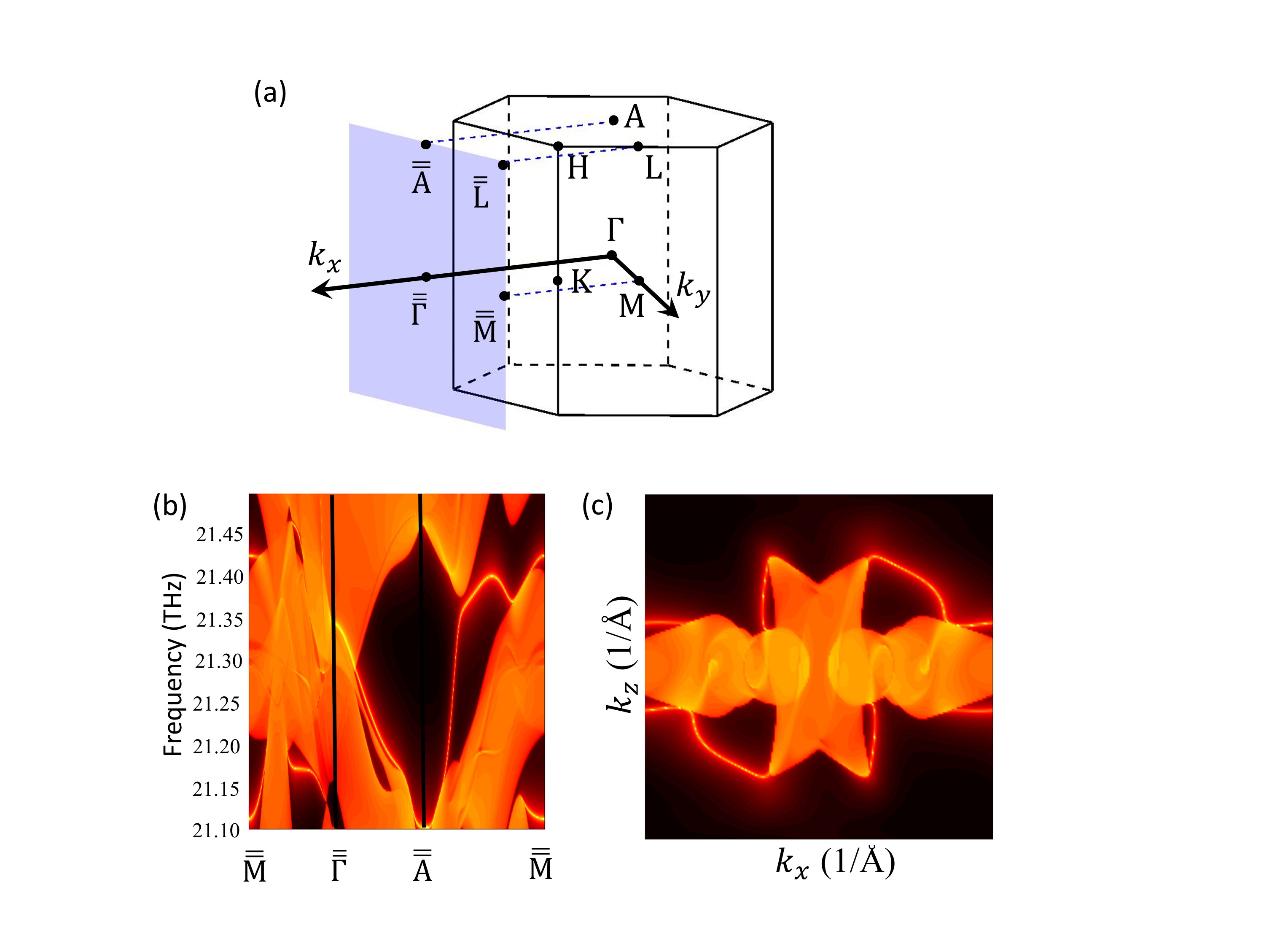}
\caption{(a)The bulk BZ and the corresponding (100) surface BZ. (b) The phonon LDOS projected on the (100) surface. (c) The isofrequency isofrequency contours of (100) surface at 21.37 THz. }
\end{figure*}

\vspace{-0.5 ex}

\renewcommand{\thetable}{S2}
\begin{table*}[h]
\centering
\setlength{\tabcolsep}{1.4mm}
\caption{ Phonon band index, positions, Chern number, and Multiplicity  of the WPs in LiIO$_3$.}
\begin{tabular}{cccccc}
\hline \hline
\   Phonon branchs & Momentum position & $\mathcal C$    &Multiplicity  \\
\hline

\     26-27 & ( 0.00000000    0.00000000    0.05243162)  & -3 & 2 \\
\   26-27& ( 0.00000000    0.00000000    -0.05243162)  & -3 &2 \\

\  26-27 &  (    0.07186095    0.16722398     0.00000000)  & 1 & 6\\
\    26-27 & ( -0.16722398    0.23908493    0.00000000)  & 1 & 6\\
\    26-27 & (   0.23908493   -0.07186095    0.00000000)  & 1 & 6\\
\     26-27 & ( -0.23908493    0.07186095    0.00000000)  & 1 & 6\\
\    26-27&  (  -0.07186095   -0.16722398   0.00000000)  & 1 & 6\\
\    26-27 & (   0.16722398   -0.23908493   0.00000000)  & 1 &6\\
\hline \hline
\end{tabular}
\end{table*}


~\\
\textbf{\large{D. The THWPs of the candidate CaTa$_4$O$_{11}$ crystallized in SG 182}}
~\\
~\\
As shown in Fig. S3(a), CaTa$_4$O$_{11}$ crystallizes in a hexagonal lattice with nonsymmorphic
space group $P6_222$ (No. 182). In our calculations, the relaxed lattice constants of CaTa$_4$O$_{11}$ are a = 6.2007 A and c = 12.3232 A,
which agree well with the experimental values a = 6.213, and c = 12.265 \cite{JAHNBERG1970454}.The hexagonal bulk BZ and its corresponding
(001) surface BZs are given in Fig. S3(b). The phonon dispersion curves along the high-symmetry directions are illustrated in Fig. S3(c).
The enlarged views of the THWP are shown in Fig. S3(d). The phonon local density of states
(LDOS) and projected isofrequency surface contours on (001) surface are shown Fig. S3(e).
We can see that the phonon surface states connect the projections of the THWP and single WPs in the
first BZ. These phonon surface states confirm the nontrivial phonon topology of THWPs in CaTa$_4$O$_{11}$.
The momentum positions and more detailed information about these nodes are presented in Table S3.

\renewcommand{\thetable}{S3}
\begin{table*}[h]
\centering
\setlength{\tabcolsep}{1.4mm}
\caption{ Phonon band index, positions, Chern number, and Multiplicity  of the WPs in CaTa$_4$O$_{11}$.}
\begin{tabular}{cccccc}
\hline \hline
\   Phonon branchs & Momentum position & $\mathcal C$    &Multiplicity  \\
\hline

\    40-41 & ( 0.00000000   0.00000000    0.33108401)  & -3 & 2 \\
\    40-41& ( 0.00000000   0.00000000    -0.33108401)  & -3 &2 \\

\    40-41 & (  0.27203327     0.27203327     0.00000000)  & 1 & 6\\
\    40-41 & (  0.27203327   -0.54406654     0.00000000)  & 1 & 6\\
\    40-41& ( -0.27203327   -0.27203327     0.00000000)  & 1 & 6\\
\    40-41 & ( -0.54406654    0.27203327    0.00000000)   & 1 & 6\\
\    40-41 & ( -0.27203327    0.54406654    0.00000000)  &  1 & 6\\
\    40-41 & ( -0.54406654   -0.27203327    0.00000000)  & 1 &  6\\

\hline \hline
\end{tabular}
\end{table*}

\renewcommand{\thefigure}{S\arabic{figure}}
\begin{figure*}
\centering
\renewcommand{\figurename}{Fig.}
\setlength{\abovecaptionskip}{-1 cm}
\includegraphics[scale=0.4]{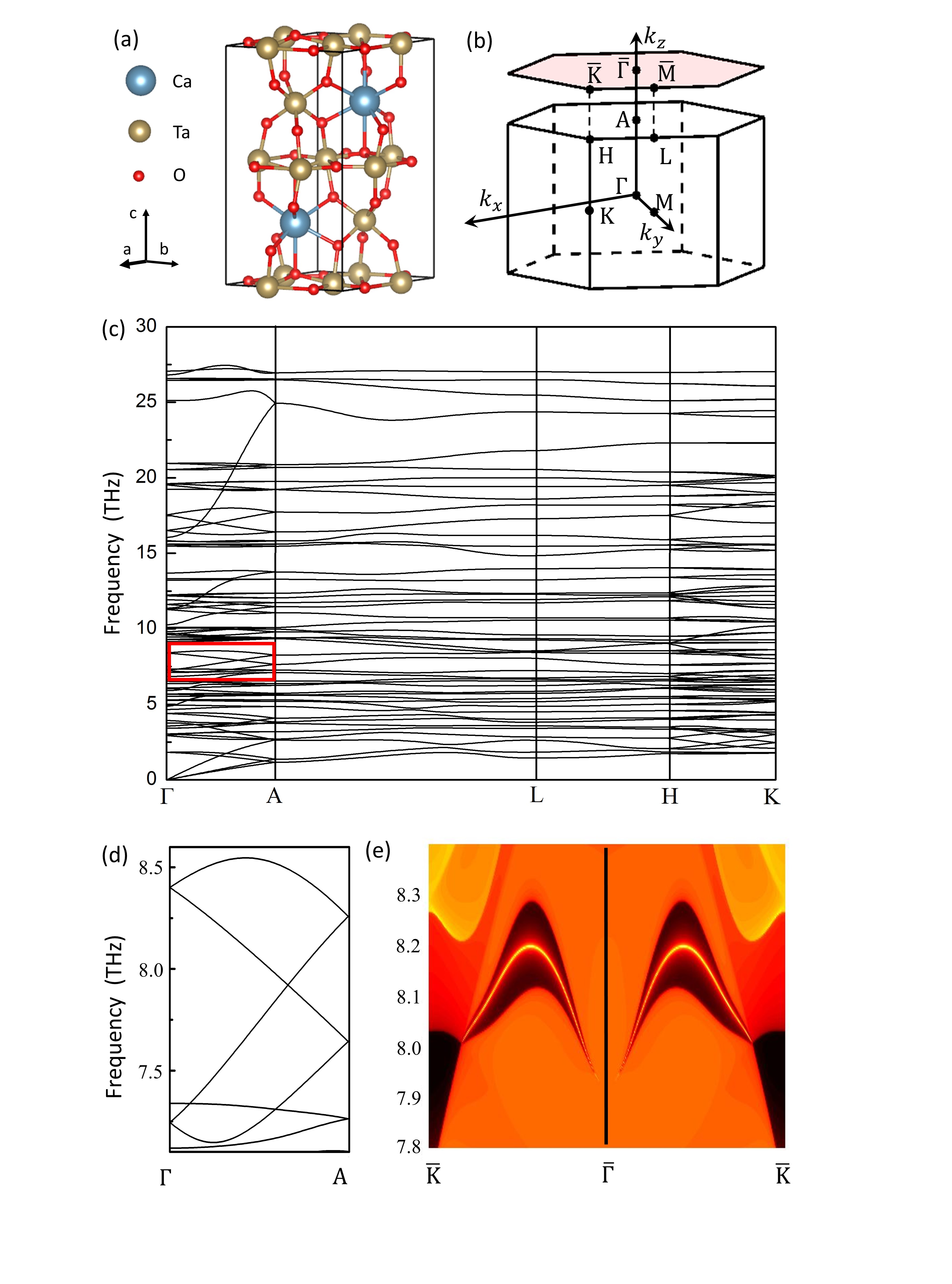}
\caption{(a) The primitive cell of CaTa$_4$O$_{11}$ with SG 182. (b) The bulk BZ and the corresponding (001) surface BZ. (c) Phonon spectrum. (d) Phonon
dispersions along $\Gamma$-A contributed from phonon branches 39 to 42, which corresponding to the red box in (c). (e) The phonon LDOS projected on the (001) surface BZ along
$\overline {\textrm K}$ - $\overline {\Gamma}$ - $\overline {\textrm K}$.  }
\end{figure*}

\end{document}